\documentclass[conference]{IEEEtran}
\IEEEoverridecommandlockouts
\usepackage{cite}
\usepackage{amsmath,amssymb,amsfonts}
\usepackage{algorithmic}
\usepackage{graphicx}
\usepackage{textcomp}
\usepackage{xcolor}
\def\BibTeX{{\rm B\kern-.05em{\sc i\kern-.025em b}\kern-.08em
    T\kern-.1667em\lower.7ex\hbox{E}\kern-.125emX}}

\usepackage{fancyhdr}
\begin{document}

\title{End-to-End Continuous Speech Emotion Recognition in Real-life Customer Service Call Center Conversations

\thanks{This project is funded by Axys Consultants and ANRT in France.}

}

\author{\IEEEauthorblockN{Yajing Feng}
\IEEEauthorblockA{\textit{LISN\textsuperscript{1,2} and AI Lab\textsuperscript{3}} \\
\textit{CNRS\textsuperscript{1}, Paris-Saclay University\textsuperscript{2} and Axys Consultants\textsuperscript{3}}\\
Orsay, France \\ 
yajing.feng@lisn.fr}
\and
\IEEEauthorblockN{Laurence Devillers}
\IEEEauthorblockA{\textit{LISN\textsuperscript{1}} \\
\textit{CNRS\textsuperscript{1}}\\
Orsay, France \\
devil@lisn.fr}
}

\maketitle
\thispagestyle{fancy}

\begin{abstract}

Speech Emotion recognition (SER) in call center conversations has emerged as a valuable tool for assessing the quality of interactions between clients and agents. In contrast to controlled laboratory environments, real-life conversations take place under uncontrolled conditions and are subject to contextual factors that influence the expression of emotions. In this paper, we present our approach to constructing a large-scale real-life dataset (CusEmo) for continuous SER in customer service call center conversations.  We adopted the dimensional emotion annotation approach to capture the subtlety, complexity, and continuity of emotions in real-life call center conversations, while annotating contextual information. The study also addresses the challenges encountered during the application of the End-to-End (E2E) SER system to the dataset, including determining the appropriate label sampling rate and input segment length, as well as integrating contextual information (interlocutor's gender and empathy level) with different weights using multi-task learning. The result shows that incorporating the empathy level information improved the model's performance.

\end{abstract}


\section{Introduction}
Building and maintaining relationships with clients is a crucial aspect of business operations, and customer service plays a vital role in this regard. The high volume of calls handled by customer service call center operators on a daily basis has spurred companies to enhance the quality of their conversations with clients, with a view to cultivating customer loyalty. To this end, the analysis of emotions in call center conversations has emerged as a valuable tool for assessing the quality of interactions between clients and agents, and ultimately improving client satisfaction \cite{anna02, wong04}, by informing the design of more personalized and adaptive service offerings. 

Automatic Speech Emotion Recognition (SER) has been extensively studied in recent decades in real-life call center conversations, with the aim of enhancing comprehension of clients' emotional states \cite{macary20, theo22, lee05}. In the previous research, the discrete emotions approach was used for modeling emotions, which is based on research on basic emotions pioneered by \cite{darwin72}, and the most commonly adopted theory within the approach is the six universal basic emotions (happiness, sadness, surprise, fear, anger, and disgust) proposed by \cite{ekman82}, which are considered as innate and not reliant on social construction. However, in real-life call center conversations, emotional variability may not always be readily apparent, and the expressions may not necessarily align with the six universal basic emotions, especially when the primary goal is information exchange. Additionally, ambivalent and ambiguous mixes of emotions may occur in the conversations\cite{feng22}. Therefore, we need an emotion modeling approach that can capture the complexity and subtle fluctuations of emotions that occur throughout the conversation. According to the dimensional approach, affective states are not independent from one another; rather, they are related to one another in a systematic manner\cite{hatice10}. The dimensional approach states that the majority of affect variability is covered by two dimensions\cite{russell80, russell83}, namely valence and arousal, which respectively represent the polarity and the intensity of the emotion, \cite{russell74} has suggested the inclusion of a third dimension, dominance, which refers the power of the sense of control over the emotion. In this research, we have adopted the dimensional approach to model latent dimensions of emotion in real-life customer service call center conversations, aiming to capture the subtlety, complexity, and continuity of emotions. The potential outcome of this research is developing a robust Continuous SER system that can provide feedback to agents or supervisors for monitoring purposes.

The subjective nature of emotional expression and perception poses a significant challenge to the field of Automatic Emotion Recognition\cite{feng22}. The diversity in social identities and personalities can result in a wide range of methods for representing emotional states. As stated by \cite{lazarus91}, we must consider the respective contributions of personality traits, knowledge, culture, environment, social structure, beliefs, goals, and expectations, in addition to the biology to the emotional process. Numerous studies addressed the question of context-sensitive affect prediction. \cite{dudzik19} provided an overview of context information (receivable elements and their knowledge and experience) in audio-visual affective databases. \cite{dudzik20} specifically discussed the influence of personal memories as a context for predicting emotional responses. \cite{andreas17} has shown the positive impact of contextual knowledge (action and relationship/environment context) on the affect-sensitive system. \cite{emotic17} explored the role of scene context information in estimating emotions from visual data.

In contrast to controlled laboratory environments, real-life call center conversations take place under uncontrolled conditions and are subject to contextual factors that influence the expression of emotions, such as the quality of the telephone communication (e.g., background noise), the purpose, resolution, and urgency of the call, and the social identity of the caller. By comprehensively understanding these contextual factors, we can gain deeper insights into emotional expressions in call center conversations. Based on this assumption, in this paper, we focus on developing a continuous SER system that incorporates the context factors in real-life customer service call center conversations. This will be achieved through a two-step process: firstly, we will create a contextual continuous emotion dataset (CusEmo Dataset), as there is currently no publicly available large-scale contextual emotional dataset of real-life customer service call center conversations. Secondly, we will discuss the application of the E2E continuous SER system in real-life customer service call center conversations and the result of incorporating the contextual factors in the system. 

In Section II, the creation of the continuous emotion dataset (CusEmo) is presented in detail, including the background of the research, data selection from large-scale real-life call center conversations, as well as the data annotation process, which includes emotional data annotation and context annotation, and the results of the data annotation. Section III presents the E2E architecture used for the continuous SER tasks in call center conversations, along with the multitask learning method used for incorporating social interaction context into the model. In Section IV, the experimental results will be discussed.

\section{CusEmo: Continuous Emotion dataset}

\subsection{Background and Corpus Design}

Our study utilized a dataset of 5548 customer service conversations obtained from a French call center in the business travel industry. The audio data was sampled at a rate of 8kHz. The conversations varied from 0.21 seconds to 25 minutes, resulting in a total duration of 375.39 hours. To facilitate the emotion annotation process, we selected a subset of customer service conversations from our larger dataset of 5548 conversations, considering the large-scale and data quality issues associated with real-life datasets. Our selection process involved four steps:

\begin{itemize}

\item \textbf{Language Selection} The call center receives calls from various regions worldwide, with conversations being conducted in multiple languages, including French, English, and Spanish. However, most of the conversations were conducted in French, with only a small proportion of conversations in other languages. Therefore, we selected only the French conversations for our initial emotional annotation task.

\item \textbf{Duration Selection} To facilitate the annotation process, we implemented a preprocessing step to delete the long silence in the audio corresponding to the information search or processing. This is done by first detecting all the silence segments in the conversation, after careful observation on these segments, we selected a threshold of 3 seconds and replaced any silence segments surpassing this duration with 1 second of silence. The choice of the threshold should be processed cautiously as silence can sometimes convey emotional information. The removed silence segments are then verified by humans before the annotation process.

\item \textbf{Speaker Diarization} The original recordings of client-agent conversations were in a single channel, making it difficult for annotators to handle the varying tones, manners, and emotional dynamics between the speakers during continuous emotion annotation. To overcome this challenge, we employed speaker diarization using SmartReport, an ASR tool based on Kaldi\cite{kaldi}. As a result, annotators could concentrate on annotating emotions for one speaker at a time, improving efficiency and accuracy.

\item \textbf{Acoustic Feature Selection} Acoustic features, including energy-related features (Shimmer, Loudness, Harmonic-to-noise Ratio), frequency-related features (Pitch, Jitter, Formants), and spectral features, have been demonstrated to reveal emotional characteristics in audio \cite{gemaps}. Specifically, previous studies have reported a high standard deviation value of F0 associated with emotions such as happiness and anger \cite{russian03}. Building on this knowledge, we utilized the OpenSMILE toolkit to extract eGEMAPS features \cite{gemaps} from each audio, and subsequently ranked the samples in descending order based on the F0 standard deviation scores. Conversations ranked higher on this scale were deemed more likely to exhibit emotional variations.

\end{itemize}

These steps allowed us to create a subset of conversations that were suitable for our emotion annotation task, while still being representative of the larger dataset. By applying these selection criteria, we aimed to reduce the noise and variability in the large-scale real-life dataset and improve the quality of the annotations.

\subsection{Continuous emotional data annotation}

\subsubsection{Background of continuous emotional datasets}

Numerous continuous emotional datasets have been developed for research purposes, including SEWA Database \cite{sewa}, MSP-Conversation corpus \cite{mspconv}, and RECOLA Database \cite{recola}. However, these datasets were either produced in a laboratory setting or in contexts other than call center conversations. To compare these existing datasets and the newly developed CusEmo Dataset, we present a detailed characteristic comparison in Table \ref{corpus}. 

\begin{table}[htbp]
\caption{Continuous emotional datasets}
\begin{center}

\begin{tabular}{|c|c|c|c|c|}
\hline
\textbf{}&\textbf{\textit{RECOLA}} & \textbf{\textit{SEWA}} & \textbf{\textit{MSP-Conv}}& \textbf{\textit{CusEmo}}\\
\hline
\textit{Environment} & Lab & Video chat& Podcast & Call center\\
\hline
\textit{Duration} & 3h50 & 15h & 44h & 15h\\ 
\hline
\textit{Language} & Fr & Multi & En & Fr\\ 
\hline
\textit{Aurio SR} & 16kHz&44.1/48kHz&16kHz&8kHz\\
\hline
\textit{label SR} & 25 & $\approx$ 66 & $\approx$ 59 & 2 \\
\hline
\textit{Value Range} & [-1,1] & [-1,1]  & [-100,100]  & [-1000,1000]  \\
\hline
\textit{Nb Annotators} & 6 & 3-5 & 2-7 & 2\\
\hline
\end{tabular}
\label{corpus}
\end{center}
\end{table}


\subsubsection{CusEmo Dataset}

In the call center conversations analyzed, we observed that the most frequently expressed categorical emotions, aside from neutral and unknown, were impatience, worry, relief, stress, gratitude, anger, confusion, and surprise. As presented previously, we adopted the dimensional approach to annotate the fluctuation of emotions throughout the conversations. However, consistently identifying a third dimension of emotion, such as dominance, other than valence and arousal, can be challenging, as noted by \cite{scherer}. Moreover, in practice, annotators often encounter difficulties in simultaneously annotating all three emotional dimensions of valence, arousal, and dominance in a 3D space for call center conversations. Thus, to mitigate these challenges, we limited our annotation to the two dimensions of valence and arousal. The 2D emotion wheel\cite{2d} was presented to the annotators as a reference for mapping categorical and dimensional emotions in a 2D space.

The annotation process for the selected conversations took place over a two-month period from July to August 2022. Two native French-speaking undergraduate psychology students, one female and one male, aged 20 and 22, were recruited as annotators for the task. In order to minimize subjectivity in the emotion annotation, additional annotators will be recruited in the subsequent phase of the study. To train the annotators for the annotation task, we provided them with a comprehensive annotation guideline, which covered the presentation of the dataset, the study objectives, discrete and dimensional emotion theories, the definition of social context information, the annotation procedure to follow, the tutorial of the annotation tool, and instructions on how their annotations would be used in this research. Additionally, we supplemented this guideline with oral instructions to ensure that the annotators fully understood the task requirements.

\begin{figure}[htbp]
\centerline{\includegraphics[width=0.6\linewidth]{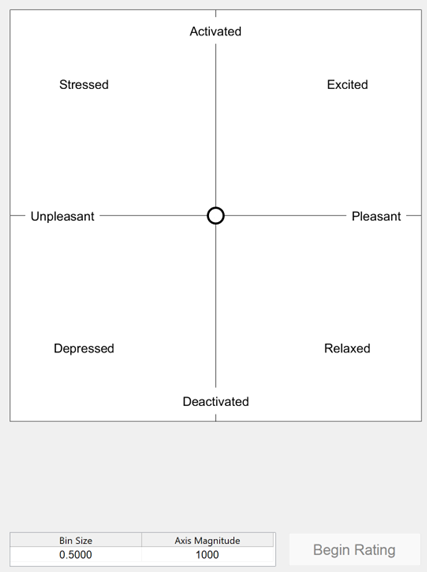}} 
\caption{Darma Annotation Tool}
\label{darma}
\end{figure}

We utilized the Darma annotation tool \cite{darma} shown in Fig.~\ref{darma} to annotate the two dimensions of valence and arousal simultaneously. We instructed annotators to move the joystick in the valence-arousal dimensional space, where the x-axis represented valence, and the y-axis represented arousal. The annotation scale ranged from -1000 to 1000 for both dimensions, where a score of 1000 represented the most positive emotional state for valence, and the most activated or intense emotional state for arousal. The annotations were done continuously, meaning that the dimensional emotional states changed constantly, and were tracked as a point continuously moving in the 2D space. The default sampling rate of Darma is 500ms, and our emotion annotations were sampled at the same rate. This resulted in two points sampled every second. 

To ensure consistency and accuracy in the annotation process, we provided annotators with several examples of emotional annotations for different audio samples. These examples were used to discuss any discrepancies or uncertainties in the annotation process, allowing for further clarification and training as needed. Furthermore, regular meetings were scheduled with the annotators to discuss any questions or issues encountered during the annotation process.

Before starting the annotation process for each audio, the annotators were instructed to verify the quality of the audio. They were asked to delete any audio files that had too much noise and where the human voice is inaudible, or consisted of waiting music or automatic responses. Furthermore, audio files with significant speaker diarization problems were also deleted. The annotation was performed separately for the conversations between the client and the agent. Given that annotating long recordings with time-continuous evaluations can be a cognitively demanding task, we instructed the annotators to take breaks between audio annotations when they felt fatigued. To minimize the potential impact of reaction time and ensure accurate alignment between the audio and emotional annotations, annotators were trained to align their annotations with the audio.

Additionally, to prevent errors or inconsistencies arising from posterior context, annotators were instructed to verify their annotations. This verification involved checking the accuracy of emotional values and timestamps. If necessary, annotators were prompted to re-annotate the conversation for improved precision. The original version (without speaker diarization) of the audio is provided to the annotators, this allows annotators to refer to the original audio whenever they have uncertainties or require additional information.

In addition to dimensional emotion annotations, we also annotated the context of conversations. We have defined two aspects of the context in the customer service call center conversations:

\begin{itemize}

    \item \textbf{Interlocutor-oriented context}: which was further categorized into social identity-related features, personality-related features, and interaction-related features. 
    
    \textbf{\textit{Social identity-related features}} were annotated categorically, including gender (male and female) as well as the social roles of the interlocutors in the conversation (individual client, secretary of the client, travel agency agent, and hotel staff, etc.).

    \textbf{\textit{Personality-related features}}, on the other hand, include empathy. \textbf{Empathy} is defined by \cite{empathy} as a complex capability enabling individuals to understand and feel the emotional states of others, resulting in compassionate behavior. We assessed the overall empathy level of interlocutors in the call center conversations based on their ability to understand and take in response the intention and emotions of others. 
    
    \textbf{\textit{Interaction-related features}}, include engagement. \textbf{Engagement} is defined by \cite{engagement} as a three-part phenomenon, which manifests in one or more of the three forms of affective, behavioral, and cognitive engagement, corresponding to the positive psychological reaction, the persistence to remain involved and the cognitively absorbed in an activity.  We evaluated the overall engagement level of the interlocutors in the conversations by their interests and willingness to actively engage in the conversation and help to resolve the problem.

    \item \textbf{Task-oriented context}: which includes the purpose (making a purchase, reporting an issue, asking for information), the resolution (resolved, unresolved, necessities client's further action, necessities agent's further action), and the urgency of the call.   

\end{itemize}

The two annotators were instructed to provide an overall evaluation of the aforementioned metadata variables related to the conversation context. The overall satisfaction level of the client is also annotated, as it can provide valuable insights into the effectiveness of call center operations. 
The urgency of the call, the empathy and the engagement of interlocutors, and the client's satisfaction is annotated with a 10-point scale, with 0 representing the least level and 10 being the strongest level. The definition of empathy and engagement along with supporting articles aforementioned has been provided to the annotators. The gender information is annotated based on human perception of the audio (voice characteristics and the content of the conversation). This involves considering factors such as the appellation used by agents towards clients (e.g., "Madam", "Sir") and the corresponding reactions from clients. In cases where the gender information cannot be discerned from the audio, it is annotated as "Unknown."

Context annotation is performed on the original conversation, encompassing the interaction between the client and the agent. This contextual annotation precedes the continuous emotion annotation process, enabling annotators to gain a comprehensive understanding of the conversation context. Annotators have the flexibility to relisten to the conversation in case of any doubts or uncertainties during the annotation process.

Currently, we have completed the context annotation for 469 conversations, as well as dimensional emotion annotation on the client's part within those conversations, and on the part of the agent in 88 conversations. These diarized conversations range in duration from 30 seconds to 10 minutes, with a total of 13.15 hours for client conversations. Furthermore, we have annotated 1.88 hours of agent conversations. In our first study, our main emphasis is on the recognition of the client's emotions. The analysis of the interaction between the interlocutors will be explored in our future studies.

\subsubsection{Analysis of annotation results}

\begin{figure}[htbp]
\centerline{\includegraphics[width=1.05\linewidth]{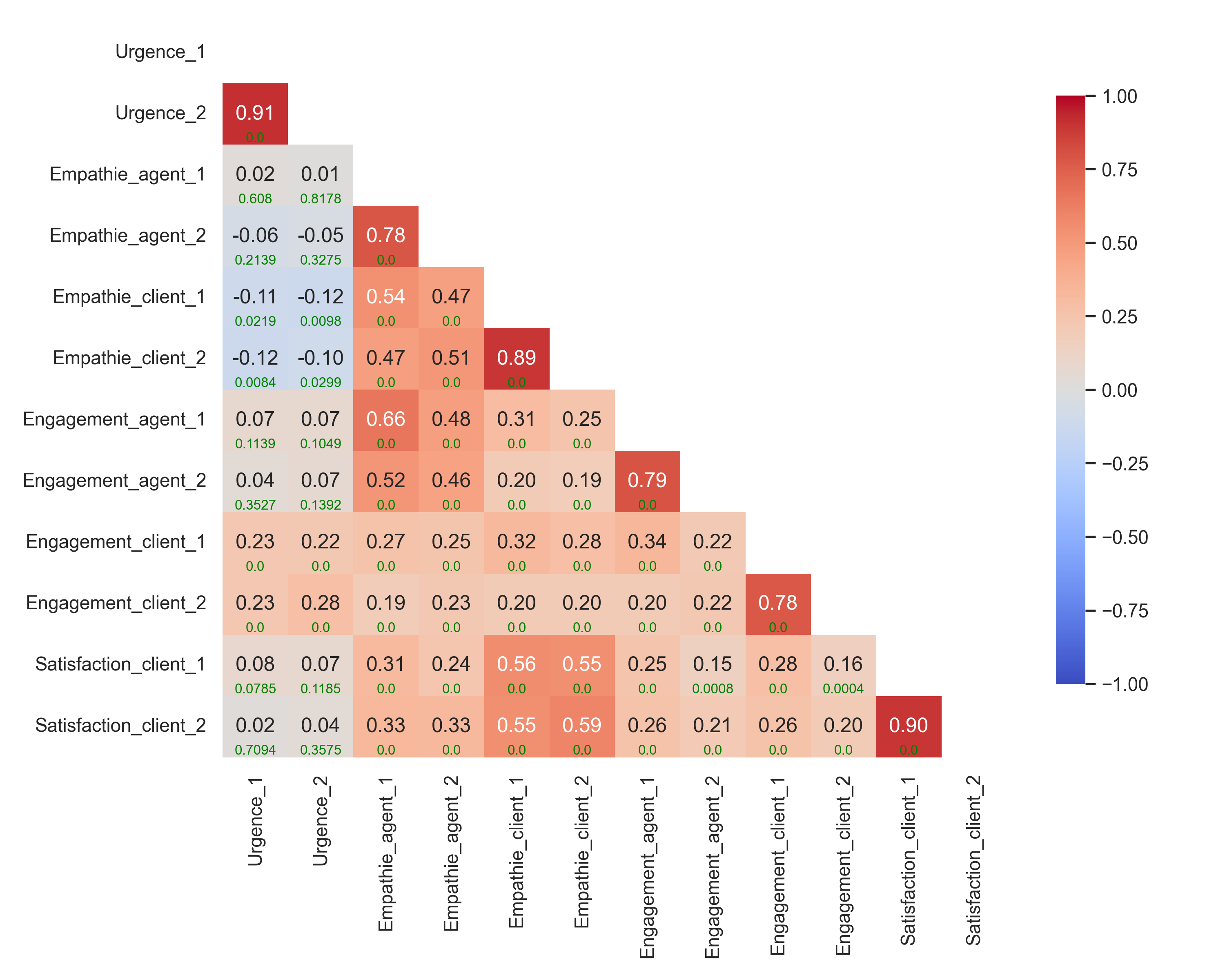}} 
\caption{Pearson correlation coefficient and statistical significance for contextual interval variables between two annotators, the number 1 (or 2) refers to annotator 1 (or 2)}
\label{pearson}
\end{figure}

In order to assess the consistency of the annotation, the inter-annotator agreement (2 annotators) has been calculated with intraclass correlation coefficient (two-way mixed, single measures), for both the valence and arousal dimensions on both the client and agent parts. The coefficients on the client's part ranged from -0.27 to 0.97 (M=0.62, SD=0.23) for valence, and from -0.65 to 0.96 (M=0.52, SD=0.28) for arousal. On the agent's part, the coefficients ranged from -0.93 to 0.97 (M=0.29, SD=0.42) for valence, and from -0.71 to 0.96 (M=0.52, SD=0.28) for arousal. Pearson correlation coefficients were calculated to examine the relationships between the contextual interval variables and between the annotations of the two annotators. The resulting coefficients and their corresponding statistical significance are presented in Figure~\ref{pearson}. It showed a high correlation between the two annotators, with Pearson correlation coefficients above 0.78 and p-values less than 0.001. Consistency and reliability of the annotations were further confirmed by assessing inter-annotator agreement using Krippendorff's alpha, which yielded similar results to Pearson correlation coefficients. We found a moderate correlation between the empathy level of the client and that of the agent. Additionally, we observed a moderate correlation between the level of empathy and the level of engagement for the agent, but this was not the case for the client. This suggests that for the client, being more engaged does not necessarily mean being more attentive to the agent's emotions, and vice versa. Furthermore, we found a moderate correlation between the empathy level of the client and their satisfaction with the conversation. This suggests that empathy may be a personality-related feature that influences the client's level of satisfaction.

\section{Model design}
\subsection{End-to-End Continuous SER}

Recognizing emotions is a challenging task due to the dynamic and nuanced nature of human emotions, which lack clear temporal boundaries\cite{e2e17multi}. Unlike the traditional handcrafted feature extraction approach\cite{gemaps} in speech-based emotion recognition (SER) tasks, which demands expertise for the manual selection of emotional features, and specifying the frame size, the End-to-End (E2E) approach automatically extracts audio features from the raw input signal and has demonstrated state-of-the-art performance in emotion recognition tasks\cite{rnn14}\cite{e2e}\cite{e2e17}\cite{e2e21}.

We adopted the E2E convolution recurrent neural network architecture proposed by \cite{e2e21b}\cite{e2e17} as Baseline. In this framework, Convolutional Neural Networks (CNN) are utilized to extract automatically paralinguistic features from raw audio.

The Long Short-Term Memory (LSTM) architecture which can retain and access long-range context information is then employed to capture the temporal information within a given context by learning the dependencies among successive inputs.

\subsection{E2E Continuous SER with Context}

To incorporate conversational context into our model, we employed a Multitask Learning (MTL) approach. This involves simultaneously training on multiple tasks, including one primary task and one or more auxiliary tasks. Previous research, such as the work by \cite{multi20}\cite{han17}\cite{multi12}\cite{multi21}, has utilized MTL to jointly learn to predict both valence and arousal dimensional emotional states. This is based on the assumption that MTL, which has more output nodes than single-task learning, can match multiple targets, and by utilizing this learning strategy, the model may improve the performance of the primary task and better predict the emotional state, as it can learn to pay more attention to samples with higher uncertainty.

The proposed network architecture consists of a Convolutional Neural Network (CNN) with three blocks, each comprising a convolution layer with 50, 125, and 125 channels and kernel sizes of 8, 6, and 6, respectively, followed by a max-pooling layer with kernel sizes of 10, 5, and 5 and the same stride size. This is succeeded by two stacked Long Short-Term Memory (LSTM) layers with a hidden layer size of 256 and a dropout rate of 0.5 to address overfitting. The output of the LSTM layers is passed through a Multilayer Perceptron (MLP) with three layers, with hidden layers of 512 and 256 neurons and an output layer having the same number of neurons as the number of predicted variables. This architecture enables the simultaneous prediction of gender and empathy level alongside the dimensional emotional states.

\subsection{Evaluation Function}

The Concordance Correlation Coefficient (CCC) is applied as the loss function to assess the model efficacy in predicting the degree of agreement between the predictions and actual labels for valence, arousal, and contextual interval variables (in our specific study the empathy level is integrated as well). As shown in 
\eqref{ccc}, $\rho$ is the Pearson correlation coefficient between X and Y, $\sigma_X$ and $\sigma_Y$ are the standard deviations of X and Y, and $\mu_X$ and $\mu_Y$ are the means of X and Y, respectively. The CCC measures the agreement between two continuous variables; it combines PCC with the square difference between the mean of the two compared time series. So as to minimize the loss, we subtract CCC from 1.

\begin{equation}
L_{CCC} = 1 - \frac{2\rho\sigma_X\sigma_Y}{\sigma_X^2+\sigma_Y^2+(\mu_X-\mu_Y)^2}\label{ccc}
\end{equation}

The Binary Cross Entropy (BCE) loss function is applied to model gender as a categorical contextual variable, which involves binary classification into Female or Male categories. As demonstrated in \eqref{bce}, BCE loss computes the discrepancy between the predicted probabilities and the true binary labels for each observation, where $N$ signifies the total number of observations in the dataset, $y_i$ denotes the actual binary label for the $i$-th instance, and $\hat{y}_i$ represents the predicted probability for the positive class. The primary objective of the BCE loss function is to estimate the probability of an input belonging to the positive class by assigning a higher loss to misclassified instances and a lower loss to correctly classified ones. The ultimate aim is to minimize the overall loss across all observations in the dataset.

\begin{equation}
L_{BCE} = - \frac{1}{N} \sum_{i=1}^{N} [y_i \log(\hat{y}_i) + (1 - y_i) \log(1 - \hat{y}_i)] \label{bce}
\end{equation}

The loss function for the model after incorporating the gender and empathy level is given by \eqref{loss}, where the respective losses for the valence (V) and arousal (A) dimensions are weighted equally. The contextual information included in the model is the Empathy level (E) and Gender (G), which are controlled by two parameters $\alpha$ and $\beta$.

\begin{equation}
L = \frac{{L_{CCC}(V) + L_{CCC}(A) + \alpha \cdot L_{CCC}(E) + \beta \cdot L_{BCE}(G)}}{{2 + \alpha + \beta}}\label{loss}
\end{equation}

\section{Experiments and results}

\subsection{Data Selection \& Ground Truth}

To ensure the reliability of the dataset, we selected a subset of 215 audio recordings of clients from 469 annotated conversations with an intraclass correlation coefficient (ICC) exceeding 0.7 on the valence dimension. For the arousal dimension, the corresponding ICC values ranged from -0.45 to 0.96 (M=0.53, SD=0.29). This resulted in a shift in the ratio of female speakers in the dataset from 2:1 to 3:1. The dataset has a total duration of 6 hours, ranging from 14 seconds to 5 minutes.

\begin{figure}[htbp]
\centerline{\includegraphics[width=\linewidth]{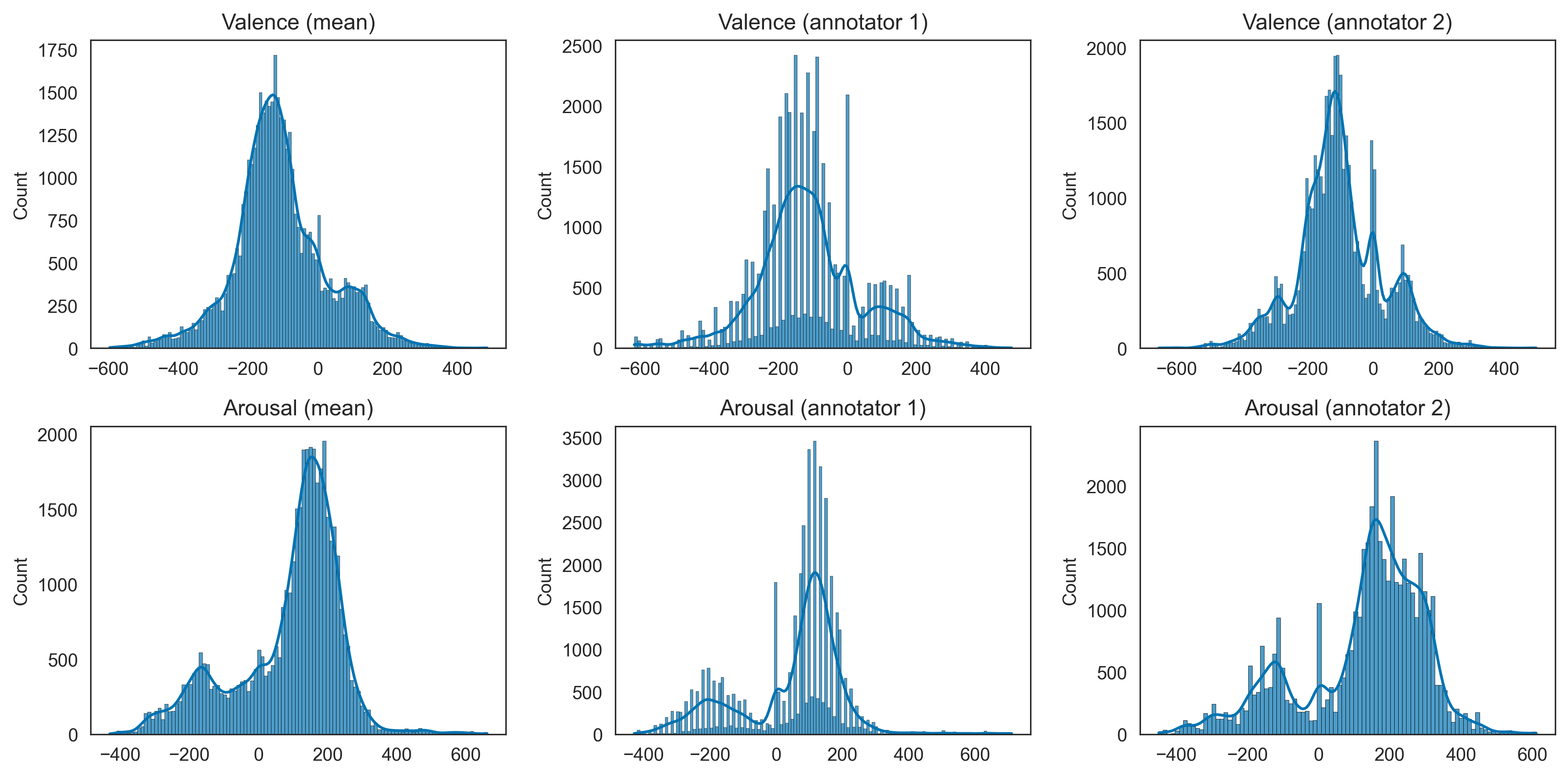}} 
\caption{Distributions of emotional attributes in the dataset.}
\label{215va}
\end{figure}

\begin{figure}[htbp]
\centerline{\includegraphics[width=\linewidth]{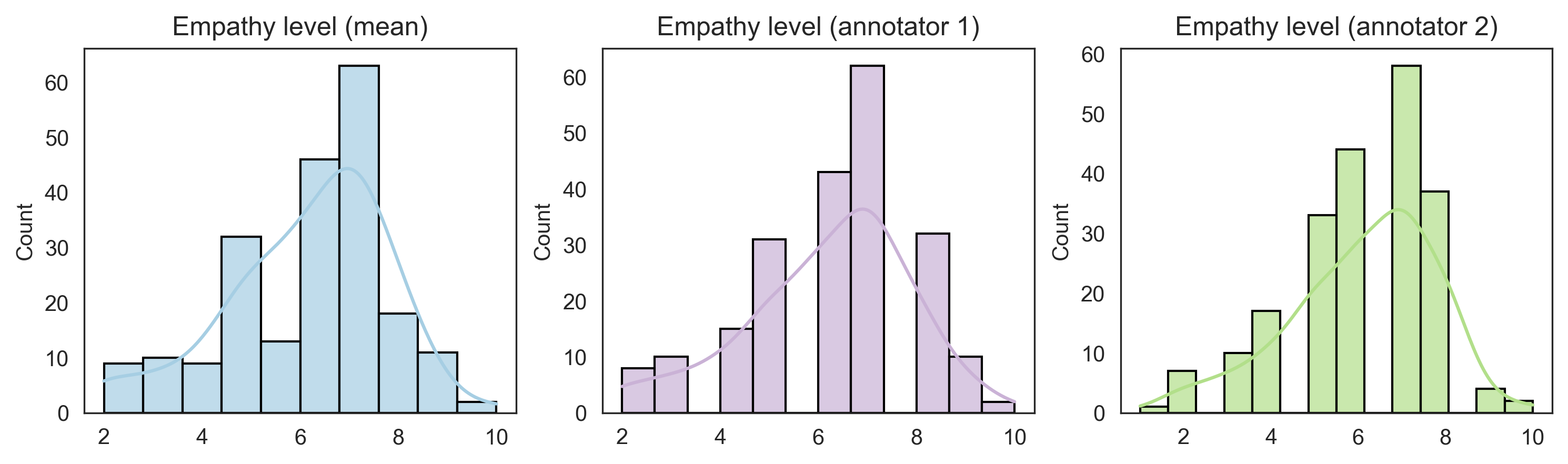}} 
\caption{Distribution of empathy level in the dataset}
\label{empathy}
\end{figure}

Figure~\ref{215va} illustrates the distribution of scores for valence and arousal dimensional annotations provided by two annotators, as well as the mean score. The score range for these annotations is -1000 to 1000. The valence annotations provided by both annotators show a distribution centered around -100, with a small peak near 100, indicating that most of the emotional annotations are slightly negative or positive. In contrast, the arousal annotations for annotator 1 are centered around 100, while for annotator 2, they are centered around 200. This suggests that annotator 2's annotations reflect a higher degree of arousal compared to annotator 1. The mean score of the two annotators is a good indicator of their annotation tendencies, as it reflects the central tendency of their annotations for both valence and arousal dimensions. Fig.~\ref{empathy} displays the score distribution of the overall evaluation of the client's empathy level (on an interval variable from 0 to 10) throughout the conversation by two annotators, based on the selected 215 audio samples. The two annotators share similar opinions on the empathy level of the client, with the mean score accurately representing this tendency.

The golden standard for the valence and arousal dimensions, as well as for the empathy level, is defined as the mean of the two annotations.

\subsection{Ablation Study}

To train our model, we utilized the Adam optimization method with a constant learning rate of $10^{-4}$ for all experiments, utilizing mini-batches of 5 samples and 150-second audio segments as the input to the E2E model. The train, development, and test sets were randomly split (with random seed 1) into sizes of 130, 45, and 40, respectively. All experiments were run six times with different random seeds (42, 24, 12, 10, 100, 125) and trained for 100 epochs using NVIDIA A100-SXM4-80GB. The duration of each training varied from 15 to 45 minutes, and the models were implemented in PyTorch.

\subsubsection{Label sampling rate}

In continuous emotion annotation tasks, the label sampling rates often vary due to the different sampling settings of various annotation tools. For instance, the MSP-Conversation Corpus and SEWA Dataset have non-uniform label sampling rates of approximately 59 and 66 frames per second, respectively. In our CusEmo dataset, the label sampling rate is 500 ms. In \cite{e2e22}, the same E2E architecture proposed by \cite{e2e21} on the MSPConv corpus, and the labels are resampled at a rate of 40ms using median filtering with a window-size of 500ms, in jumps of 1/59s. To investigate whether the label sampling rate would impact the result, we resampled our dataset at different rates using Cubic spline interpolation.

\begin{table}[htbp]
\caption{Sampling rate comparison}
\begin{center}
\begin{tabular}{|c|c|c|}
\hline
\textbf{} & \textbf{\textit{Valence}} & \textbf{\textit{Arousal}} \\
\hline
\textit{sr=500ms} & \textbf{.667} ($ .044 $)& \textbf{.499} ($ .014$) \\
\hline
\textit{sr=200ms} & .516 ($ .022 $)& .427 ($ .004$) \\
\hline
\textit{sr=100ms} & .208 ($ .003 $)& .212 ($ .007$) \\
\hline
\textit{sr=40ms} & .078 ($ .003$) & .093 ($ .001$) \\ 
\hline
\end{tabular}
\label{sr}
\end{center}
\end{table}

The results presented in Table \ref{sr} indicate a gradual decrease in model performance as the sampling rate decreases from 500ms to 40ms. Specifically, the mean scores of CCC(v) and CCC(a) decrease from 0.657 and 0.434 with standard deviations of 0.022 and 0.003 to 0.004 and 0.093 with standard deviations of 0.004 and 0.001. This may be due to the fact that there is not a large variation in emotional states of the speaker during the call center conversations. Therefore, increasing the number of annotation points may not necessarily improve the model's ability to learn.

The model was also tested on the MSP-Conv Corpus, with the label sampling rate changed from the non-uniform 59 fps to a fixed rate of 40ms. The resulting CCC(v) and CCC(a) were 0.19 and 0.3, respectively, which are similar to the results reported in \cite{e2e22}. However, it should be noted that our dataset consists of diarized audio containing only the client's speech, unlike the MSP-Conv Corpus which is a non-diarized podcast recording. Therefore, it is difficult to make direct comparisons between the results obtained from the two datasets.

\subsubsection{Batch size \& Segment length}

Given the lack of clear temporal boundaries for human emotions, we aimed to explore the optimal input segment size for our E2E model. To investigate this, we imposed a limit on the total audio duration in each batch, and examined whether varying the input segment size had an impact on the model's performance.

\begin{table}[htbp]
\caption{Segment size comparison}
\begin{center}
\begin{tabular}{|c|c|c|c|c|}
\hline
\textbf{\textit{Segment size}} & \textbf{\textit{Batch size}} & \textbf{\textit{Valence}} & \textbf{\textit{Arousal}} \\
\hline
\textit{300s} & \textit{5} & \textbf{.667} ($ .044$)& \textbf{.499} ($ .014$) \\
\hline
\textit{150s} & \textit{10} & \textbf{.667} ($ .039$)& .475 ($ .055$) \\
\hline
\textit{100s} & \textit{15} & .642 ($ .033$)& .486 ($ .02$) \\
\hline
\textit{75s} & \textit{20} & .64 ($ .026$)& .486 ($ .013$) \\
\hline
\textit{50s} & \textit{30} & .634 ($ .012$)& \textbf{.499} ($ .008$) \\
\hline
\end{tabular}
\label{seg_size}
\end{center}
\end{table}

The results presented in Table \ref{seg_size} indicate that increasing the segment size can lead to an overall increase in the CCC(v). However, changing the segment size does not appear to have a significant impact on the CCC(a). As the highest CCC(a) score is achieved when the segment size is set to both 50 seconds and 300 seconds.

\subsubsection{Contextual learning: gender and empathy level}

In this study, our aim was to investigate whether incorporating contextual information into the model could provide additional information and subsequently improve the model's performance. To achieve this, we first examined two variables providing the interlocutor-oriented context information: gender as a categorical variable (Male or Female), and empathy as an interval variable (noted as an integer between 0 and 10). Additionally, different weights for integrating the contextual information in the multitask learning model were tested, for a segment size of 300s and batch size of 5, and the results are presented in Table \ref{context}.

\begin{table}[htbp]
\caption{contextual information weights comparison}
\begin{center}
\begin{tabular}{|c|c|c|c|c|}
\hline
\textbf{\textit{$\alpha$}} & \textbf{\textit{$\beta$}} & \textbf{\textit{Valence}} & \textbf{\textit{Arousal}} \\
\hline
\textit{0} & \textit{0} & \textbf{.667} ($ .044$)& \textbf{.499} ($ .014$) \\
\hline
\textit{0} & \textit{1} & .213 ($ .011$)& .141 ($ .009$) \\
\hline
\textit{1} & \textit{0} & .649 ($ .05$)& .492 ($ .018$) \\
\hline
\textit{0.1} & \textit{0} & \textbf{.699} ($ .014$)& .497 ($ .009$) \\
\hline
\textit{0.01} & \textit{0} & .676 ($ .025$)& .502 ($ .012$) \\
\hline
\textit{0.001} & \textit{0} & \textbf{.684} ($ .012$)& \textbf{.508} ($ .013$) \\
\hline

\textit{0} & \textit{0.1} & .565 ($ .175$)& .435 ($ .154$) \\
\hline
\textit{0} & \textit{0.01} & .652 ($ .034$)& .494 ($ .027$) \\
\hline
\textit{0} & \textit{0.001} & \textbf{.667} ($ .018$)& .491 ($ .003$) \\
\hline
\end{tabular}
\label{context}
\end{center}
\end{table}

We observed that incorporating gender or empathy level information with equal weights for valence and arousal did not improve model performance. However, when incorporating the empathy level with a weight of 0.1, the best CCC score was achieved for the valence dimension. To determine the statistical significance of this result, a Wilcoxon signed-rank test was conducted, resulting in a p-value of 0.01, which indicates a significant outcome. For the arousal dimension, the best CCC score was obtained when incorporating empathy level with a weight of 0.001, and the p-value obtained is 0.015, indicating a significant outcome. Upon incorporating gender information, we did not observe any improvement in the results. The best outcome for the valence dimension was achieved when using a weight of 0.001. However, the obtained p-value of 0.71 suggests that the observed differences in the results are likely due to random chance. Furthermore, gender is a categorical variable, empathy is an interval variable, and valence and arousal are two ratio variables. The approach of incorporating gender information along with valence, arousal, and empathy level in the multitask learning approach requires further investigation. In addition to exploring limited values of {$\alpha$} and {$\beta$}, we also tested autoML approach with Optuna to determine the optimal weight for empathy, we conducted an experiment (random seed 12, 60 training epochs) using TPE (Tree-structured Parzen Estimator), a Bayesian optimization algorithm based on kernel fitting, the number of trials conducted was set to 10. The best result was obtained with {$\alpha$} equal to 0.88, resulting in a CCC score of 0.67 for valence and 0.43 for arousal. However, to enhance the robustness and reliability of the obtained results, we intend to conduct additional experiments with a larger number of trials.

\section{Conclusions and Future work}

In this study, we presented the application of the E2E SER system in real-life customer service call center conversations. We introduced our approach for building a large-scale real-life dataset that adopts the dimensional emotion approach for continuous emotion annotation in call center conversations, while including annotations of the contextual information. Additionally, we addressed the challenges we encountered during the application of the E2E SER system to our dataset, including determining the appropriate label sampling rate, and segment length, and integrating contextual information in the model with different weights, to enhance the model performance on valence and arousal prediction. By adopting multitask learning approach to incorporating the context information, we conclude that the integration of the empathy level information can improve the model's performance. In the next steps of our research, we will explore further approaches to incorporate contextual information. Our plan also involves the development of emotion recognition systems that leverage both automatic text transcription and speech data. Additionally, we will place particular emphasis on the detection of significant emotional changes during conversations.



\section*{Ethical Impact Statement}

The customer service call center conversations are obtained from a French call center in the business travel industry, between August 26th and September 27th, 2021. Interlocutors provided consent for the recording and the utilization of the conversations to enhance service quality. To comply with GDPR regulations, personal information such as names, phone numbers, emails, and addresses were anonymized. In the audio recordings, personal information was replaced with white noise of equal duration, while in the speech transcriptions, named entities were used. The dataset is utilized for internal research and development purposes within the company, no convention is signed to publish the dataset at the current stage.

\section*{Acknowledgment}

Yajing Feng thanks her co-supervisors Jean-Marc Guidicelli and Laurence Suprano at Axys Consultants for their invaluable guidance and support throughout the research process, as well as her colleague François Buet at LISN-CNRS for providing technical support and feedback on the manuscript.

\end{document}